\documentclass[prd,a4paper,twocolumn]{revtex4}
\usepackage{graphicx}
\usepackage[]{subfigure}
\usepackage{epsfig}

\newcommand{\nc}{\newcommand}

\nc{\be}[1]{\begin{equation}\mbox{$\label{#1}$}}
\nc{\bea}[1]{\begin{eqnarray} \mbox{$\label{#1}$}}
\nc{\Section}[2]{\section{#2}\label{#1}}
\nc{\Bibitem}[1]{\bibitem{#1}}
\nc{\Label}[1]{\label{#1}}

\nc{\eea}{\end{eqnarray}}
\nc{\ee}{\end{equation}}

\nc{\bdm}{\begin{displaymath}}
\nc{\edm}{\end{displaymath}}
\nc{\dpsty}{\displaystyle}
\nc{\bc}{\begin{center}}
\nc{\ec}{\end{center}}
\nc{\ba}{\begin{array}}
\nc{\ea}{\end{array}}
\nc{\bab}{\begin{abstract}}
\nc{\eab}{\end{abstract}}
\nc{\btab}{\begin{tabular}}
\nc{\etab}{\end{tabular}}
\nc{\bit}{\begin{itemize}}
\nc{\eit}{\end{itemize}}
\nc{\ben}{\begin{enumerate}}
\nc{\een}{\end{enumerate}}
\nc{\bfig}{\begin{figure}}
\nc{\efig}{\end{figure}}

\nc{\arreq}{&\!=\!&}
\nc{\arrmi}{&\!-\!&}
\nc{\arrpl}{&\!+\!&}
\nc{\arrap}{&\!\!\!\approx\!\!\!&}
\nc{\non}{\nonumber}
\nc{\align}{\!\!\!\!\!\!\!\!&&}

\def\lsim{\; \raise0.3ex\hbox{$<$\kern-0.75em
      \raise-1.1ex\hbox{$\sim$}}\; }
\def\gsim{\; \raise0.3ex\hbox{$>$\kern-0.75em
      \raise-1.1ex\hbox{$\sim$}}\; }

\nc{\DOT}{\hspace{-0.08in}{\bf .}\hspace{0.1in}}
\nc{\Laada}{\hbox {$\sqcap$ \kern -1em $\sqcup$}}
\nc\loota{{\scriptstyle\sqcap\kern-0.55em\hbox{$\scriptstyle\sqcup$}}}
\nc\Loota{{\sqcap\kern-0.65em\hbox{$\sqcup$}}}
\nc\laada{\Loota}
\nc{\qed}{\hskip 3em \hbox{\BOX} \vskip 2ex}

\nc{\real}{{\rm I \! R}}
\nc{\Z}{{\sf Z \!\!\! Z}}
\nc{\complex}{{\rm C\!\!\! {\sf I}\,\,}}
\def\bigid{\leavevmode\hbox{\small1\kern-3.8pt\normalsize1}}
\def\id{\leavevmode\hbox{\small1\kern-3.3pt\normalsize1}}
\nc{\slask}{\!\!\!/}
\nc{\bis}{{\prime\prime}}
\nc{\pa}{\partial}
\nc{\na}{\nabla}
\nc{\ra}{\rangle}
\nc{\la}{\langle}
\nc{\goto}{\rightarrow}
\nc{\swap}{\leftrightarrow}

\nc{\EE}[1]{ \mbox{$\cdot10^{#1}$} }
\nc{\abs}[1]{\left|#1\right|}
\nc{\at}[2]{\left.#1\right|_{#2}}
\nc{\norm}[1]{\|#1\|}
\nc{\abscut}[2]{\Abs{#1}_{\scriptscriptstyle#2}}
\nc{\vek}[1]{{\rm\bf #1}}
\nc{\integral}[2]{\int\limits_{#1}^{#2}}
\nc{\inv}[1]{\frac{1}{#1}}
\nc{\dd}[2]{{{\partial #1}\over{\partial #2}}}
\nc{\ddd}[2]{{{{\partial}^2 #1}\over{\partial {#2}^2}}}
\nc{\dddd}[3]{{{{\partial}^2 #1}\over
    {\partial #2 \partial #3}}}
\nc{\dder}[2]{{{d #1}\over{d #2}}}
\nc{\ddder}[2]{{{d^2 #1}\over{d {#2}^2}}}
\nc{\dddder}[3]{{d^2 #1}\over
    {d #2 d #3}}
\nc{\dx}[1]{d\,^{#1}x}
\nc{\dy}[1]{d\,^{#1}y}
\nc{\dz}[1]{d\,^{#1}z}
\nc{\dl}[1]{\frac{d\,^{#1}l}{(2\pi)^{#1}}}
\nc{\dk}[1]{\frac{d\,^{#1}k}{(2\pi)^{#1}}}
\nc{\dq}[1]{\frac{d\,^{#1}q}{(2\pi)^{#1}}}

\nc{\bfT}{{\bf T }}

\nc{\cA}{{\cal A}}
\nc{\cB}{{\cal B}}
\nc{\cD}{{\cal D}}
\nc{\cE}{{\cal E}}
\nc{\cG}{{\cal G}}
\nc{\cH}{{\cal H}}
\nc{\cL}{{\cal L}}
\nc{\cO}{{\cal O}}
\nc{\cT}{{\cal T}}
\nc{\cN}{{\cal N}}
\nc{\cR}{{\cal R}}

\nc{\rvac}[1]{|{\cal O}#1\rangle}
\nc{\lvac}[1]{\langle{\cal O}#1|}
\nc{\rvacb}[1]{|{\cal O}_\beta #1\rangle}
\nc{\lvacb}[1]{\langle{\cal O}_\beta #1 |}
\nc{\bb}{\bar{\beta}}
\nc{\bt}{\tilde{\beta}}
\nc{\ctH}{\tilde{\cal H}}
\nc{\chH}{\hat{\cal H}}
\nc{\1}{\aa}
\nc{\2}{\"{a}}
\nc{\3}{\"{o}}
\nc{\4}{\AA}
\nc{\5}{\"{A}}
\nc{\6}{\"{O}}
\nc{\al}{\alpha}
\nc{\g}{\gamma}
\nc{\Del}{\Delta}
\nc{\e}{\textrm{e}}
\nc{\eps}{\epsilon}
\nc{\lam}{\lambda}
\nc{\Om}{\Omega}
\nc{\ve}{\varepsilon}
\nc{\mn}{{\mu\nu}}
\nc{\vp}{\varphi}

\nc{\rf}[1]{(\ref{#1})}
\nc{\nn}{\nonumber \\*}
\nc{\bfB}{\bf{B}}
\nc{\bfv}{\bf{v}}
\nc{\bfx}{\bf{x}}
\nc{\bfy}{\bf{y}}
\nc{\vx}{\vec{x}}
\nc{\vy}{\vec{y}}
\nc{\oB}{\overline{B}}
\nc{\oI}{\overline{I}}
\nc{\oR}{\overline{R}}
\nc{\rar}{\rightarrow}
\nc{\ti}{\times}
\nc{\slsh}{\hskip-5pt/}
\nc{\sm}{Standard~Model~}
\nc{\MP}{M_{\rm Pl}}
\nc{\mpl}{M_{\rm Pl}}
\nc{\tp}{t_{\rm Pl}}

\nc{\pmin}{p_{\rm min}}
\nc{\pmax}{p_{\rm max}}
\nc{\fo}{f_0}
\nc{\foi}{f_{0,i}\,}
\nc{\fop}{f_0^P}
\nc{\fou}{f_0^U}

\nc{\eff}{{\rm eff}}
\nc{\MT}{M_{\rm T}}
\nc{\ML}{M_{\rm L}}
\nc{\kk}{\vek{k}}
\nc{\pp}{{\rm p}}
\nc{\pt}{\partial_t}
\nc{\half}{{1\over 2}}
\nc{\w}{\omega}
\nc{\uhat}{\hat{U}_\w}

\nc{\etal}{\mbox{\it et al.}}
\nc{\ie}{{\it i.e. }}
\nc{\eg}{{\it e.g. }}
\nc{\trh}{T_{\rm RH}}
\nc{\ad}{{a'\over a}}
\nc{\bd}{{b'\over b}}
\nc{\Rd}{{R'\over R}}
\nc{\diag}{{\textrm{diag}}}
\nc{\mato}[1]{\tilde{#1}}
\nc{\sinn}{\textrm{sinn}}
\nc{\sech}{\textrm{sech}}
\nc{\I}{\textrm{I}}
\nc{\II}{\textrm{II}}
\nc{\III}{\textrm{III}}
\nc{\vev}[1]{\langle #1 \rangle}
\nc{\hyp}{\,\; F_{1{\hskip -16pt}2}{\hskip 11pt}}
\nc{\brhom}{\overline{\rho}_M}
\nc{\brho}{\overline{\rho}}
\nc{\rhob}{\overline{\rho}}
\nc{\Pb}{\overline{P}}
\nc{\bH}{\overline{H}}
\nc{\ep}{{1+4\eps}}

\nc{\deriv}[2]{ 
\frac{\mathrm{d}#1}{\mathrm{d}#2}
}
\nc{\Mnu}{M_\nu}
\nc{\bee}{\begin{equation}}
\nc{\ene}{\end{equation}}
\nc{\hdp}{\sigma_8 (\Omega_{\rm m}/0.3)^{0.37}}

\def\smiley{\hbox{\large$\bigcirc$\hspace{-.80em}%
\raise.2ex\hbox{$\cdot\cdot$}\kern-.61em    
\lower.2ex\hbox{\scriptsize$\smile$}}\ }

\def\frowney{\hbox{\large$\bigcirc$\hspace{-.80em}%
\raise.2ex\hbox{$\cdot\cdot$}\kern-.635em
\lower.2ex\hbox{\scriptsize$\frown$}}\ }

\begin{document}

\title{Using the cluster mass function from weak lensing to constrain neutrino masses}
\author{Jostein R. Kristiansen}
\email{j.r.kristiansen@astro.uio.no}
\affiliation{Institute of Theoretical Astrophysics, University of
  Oslo, Box 1029, 0315 Oslo, NORWAY}
\author{{\O}ystein Elgar\o y}
\email{oelgaroy@astro.uio.no}
\affiliation{Institute of Theoretical Astrophysics, University of
  Oslo, Box 1029, 0315 Oslo, NORWAY}
\author{H\aa kon Dahle}
\email{hakon.dahle@astro.uio.no}
\affiliation{Institute of Theoretical Astrophysics, University of
  Oslo, Box 1029, 0315 Oslo, NORWAY}
\date{\today}

\begin{abstract}
We discuss the variation of cosmological 
upper bounds on $M_\nu$, the sum of the neutrino masses, 
with the choice of data sets included in the 
analysis, pointing out a few oddities not easily seen 
when all data sets are combined. For example, the effect of
applying different priors varies significantly depending on whether we
use the power spectrum from the 2dFGRS or SDSS galaxy survey.   
A conservative neutrino mass limit of $M_\nu < 1.43$eV (95\%C.L.)  
is obtained by combining the 
WMAP 3 year data with the cluster mass function measured by 
weak gravitational lensing.  This limit has the virtue of not making 
any assumptions about the bias of luminous matter with respect to 
the dark matter, and is in this sense (and this sense only) bias-free. 
\end{abstract}

\maketitle

\section{Introduction}   

The fact that neutrinos undergo flavour oscillations, 
implying that not all neutrinos 
are massless, is one of the most important discoveries in particle physics in 
the last decade.  Oscillations are only sensitive to the mass-squared differences between the neutrino mass eigenstates (see e.g. \cite{maltoni:2004,fogli:2006}), 
and although these are suggestive 
of the overall mass scale, they cannot pin the absolute mass scale firmly.  
For that, one needs probes like tritium beta decay \cite{eitel:2005}
or neutrinoless double 
beta decay \cite{elliott:2002}.  
At the moment, however, the strongest upper bound on the 
neutrino mass scale comes from cosmology (see e.g. \cite{goobar:2006,seljak:2006}).  
The cosmological mass limits 
are based on the implications of neutrino masses for clustering of 
matter, in particular the fact that neutrinos can free-stream out of 
density perturbations and impede structure formation on small scales 
if they are a significant fraction of the dark matter.  For a 
thorough review, see \cite{lesgourgues:2006}. 

Impressive as they are, the cosmological neutrino mass limits involve 
several assumptions.  First of all, they assume that the underlying 
cosmological model is of the Friedmann-Robertson-Walker type, that 
gravity is described by general relativity, the primordial power spectrum 
of density perturbations is a scale-free power-law, and that the 
dark energy is a cosmological constant.  Modifications of this simple 
picture have been considered \cite{zunckel:2006}, and in particular the degeneracy 
between neutrino masses 
and the dark energy equation of state has been investigated in some recent 
papers \cite{hannestad:2005,goobar:2006,ichikawa:2005b,delamacorra:2006,xia:2006}.  
Secondly, the upper limit depends on the cosmological data 
sets used in the analysis.   
It is this last aspect of the problem we will 
consider in the present paper, working within the context of the 
standard $\Lambda$CDM paradigm.   

The angular power spectrum of the temperature anisotropies in the 
cosmic microwave background (CMB) radiation is generally considered to be 
the cleanest cosmological 
data set.  Unfortunately the CMB temperature power spectrum cannot provide upper limits on $\Mnu$ better 
than $\sim 1.6$ eV \cite{ichikawa:2005}, which is almost reached already 
now with the WMAP 3-year data \cite{spergel:2006,fukugita:2006,kristiansen:2006}.
An improvement of this limit requires some probe of the 
matter distribution.  The most common probe is the power spectrum 
of the galaxy distribution, as determined by large redshift surveys 
like the 2 degree Field Galaxy Redshift Survey (2dFGRS) \cite{cole:2005} 
and the 
Sloan Digital Sky Survey (SDSS) \cite{tegmark:2003}.  The distribution of the galaxies 
is biased with respect to that of the dark matter, but if the bias 
is independent of length scale, this is not a serious limitation.  
However, recent results suggest that the story of bias might be 
more complicated than the simple assumption of scale-independent 
bias that has gone into earlier cosmological mass limits 
\cite{percival:2006,smith:2006}.    

Ideally, one would like to rely on probes that are sensitive to the 
total matter distribution.  One such probe is measurements of 
weak gravitational lensing by matter along random lines of sight 
\cite{refregier:2003, schneider:2006}. However, even the most ambitious 
of these 'cosmic shear' surveys have covered less than one percent of the full sky, 
thus probing only a limited cosmological volume.
A somewhat different approach is to count the number of rare, high density 
peaks in the matter fluctuations within a larger volume. 
These peaks correspond to very massive clusters of galaxies, their abundances 
(the so-called cluster mass function) 
being highly sensitive to the amplitude of the matter power spectrum 
(see e.g.\ \cite{voit:2005} and references therein) on corresponding scales. 
Until recently, cluster masses were generally
derived based on observables of the baryonic mass component, 
such as the temperature of the X-ray emitting intra-cluster medium (ICM). In this case, 
one had to calibrate the X-ray temperature-mass relationship, either 
from simulations or from observational mass measurements relying on 
assumptions about hydrostatical equilibrium of the ICM.  
However, Dahle \cite{dahle:2006} provided for the first time a measurement of the cluster mass function (CMF) 
where the 
cluster masses have been established directly by weak gravitational lensing.  
The main result of this paper is an upper bound on 
the sum of the neutrino 
masses obtained by combining the WMAP 3-year data with the CMF 
derived by Dahle \cite{dahle:2006}.  Although not as impressive as e.g. the limit obtained by 
combining all available cosmological observations, including the Lyman 
alpha forest power spectrum (see e.g. \cite{seljak:2006, cirelli:2006}), this limit is robust in the sense that 
it makes use of clean cosmological probes, and involves a minimum of 
assumptions.  

\section{Massive neutrinos in cosmology}

We work within the standard cosmological paradigm of the $\Lambda$CDM 
model, and leave studies of significant deviations from this model 
for a future study. Our adopted notation is as follows: We denote as 
$\Omega_{\rm i}$ the present density of component ${\rm i}$ in units of the 
density of a spatially flat universe.  Note that $\Omega_{\rm m}$ is 
the total density of all non-relativistic components, so that 
$\Omega_{\rm m} = \Omega_{\rm{CDM}} +\Omega_{\rm b}+\Omega_\nu$, with $\Omega_{\rm b}$ 
being the present baryon density.  The density parameter $\Omega_\nu$ 
is determined by the sum of the neutrino masses $M_\nu$ and the 
present value of the Hubble parameter $H_0=100 h\;{\rm km}\,{\rm s}^{-1}
\,{\rm Mpc}^{-1}$ \cite{mangano:2005}:
\begin{equation}
\Omega_\nu h ^2 = \frac{M_\nu}{93.14\;{\rm eV}}.
\label{eq:omeganu}
\end{equation}
As we will see, present cosmological probes (with the possible exception 
of the Lyman alpha forest \cite{seljak:2006}) are only sensitive to 
neutrino masses larger 
than a few tenths of an eV.  In this regime, the neutrino mass spectrum 
would be degenerate, and hence $M_\nu = 3 m_\nu$, where $m_\nu$ is the 
common mass of the neutrino mass eigenstates. 

Within the $\Lambda$CDM model the effect 
of massive neutrinos on structure formation is simple.  
Their influence on structure formation is governed mainly by the 
quantity $f_\nu = \Omega_\nu / \Omega_{\rm m}$. While relativistic, 
the neutrinos free-stream out of density perturbations on scales 
smaller than the Hubble radius.  Once they become non-relativistic, 
the free-streaming scale is set by their root-mean-square velocity.  
The maximum length scale below which all scales are affected by 
neutrino free-streaming is thus set by the horizon size when the 
neutrinos became non-relativistic.  This quantity is given by \cite{lesgourgues:2006}
\begin{eqnarray}
k_{\rm nr} &=& 0.018\left(\frac{m_\nu}{1\;{\rm eV}}\right)^{1/2}
\Omega_{\rm m}^{1/2}\;h\,{\rm Mpc}^{-1} \nonumber \\  
&=& 0.1f_\nu^{1/2}\Omega_{\rm m}h\;h\,{\rm Mpc}^{-1}.
\label{eq:knr}
\end{eqnarray}
Note that the combination $\Omega_{\rm m}h$ enters this expression; 
this quantity determines the Hubble radius at matter-radiation equality 
and is an important length scale in the power spectrum of matter density 
fluctuations.  This scale is similar to the neutrino free-streaming 
scale: below this scale, density perturbations are suppressed. 
Hence, there is a degeneracy between $\Omega_{\rm m} h$ 
and $f_\nu$.  

\begin{figure}
\includegraphics[width=4cm,height=4cm,angle=-90]{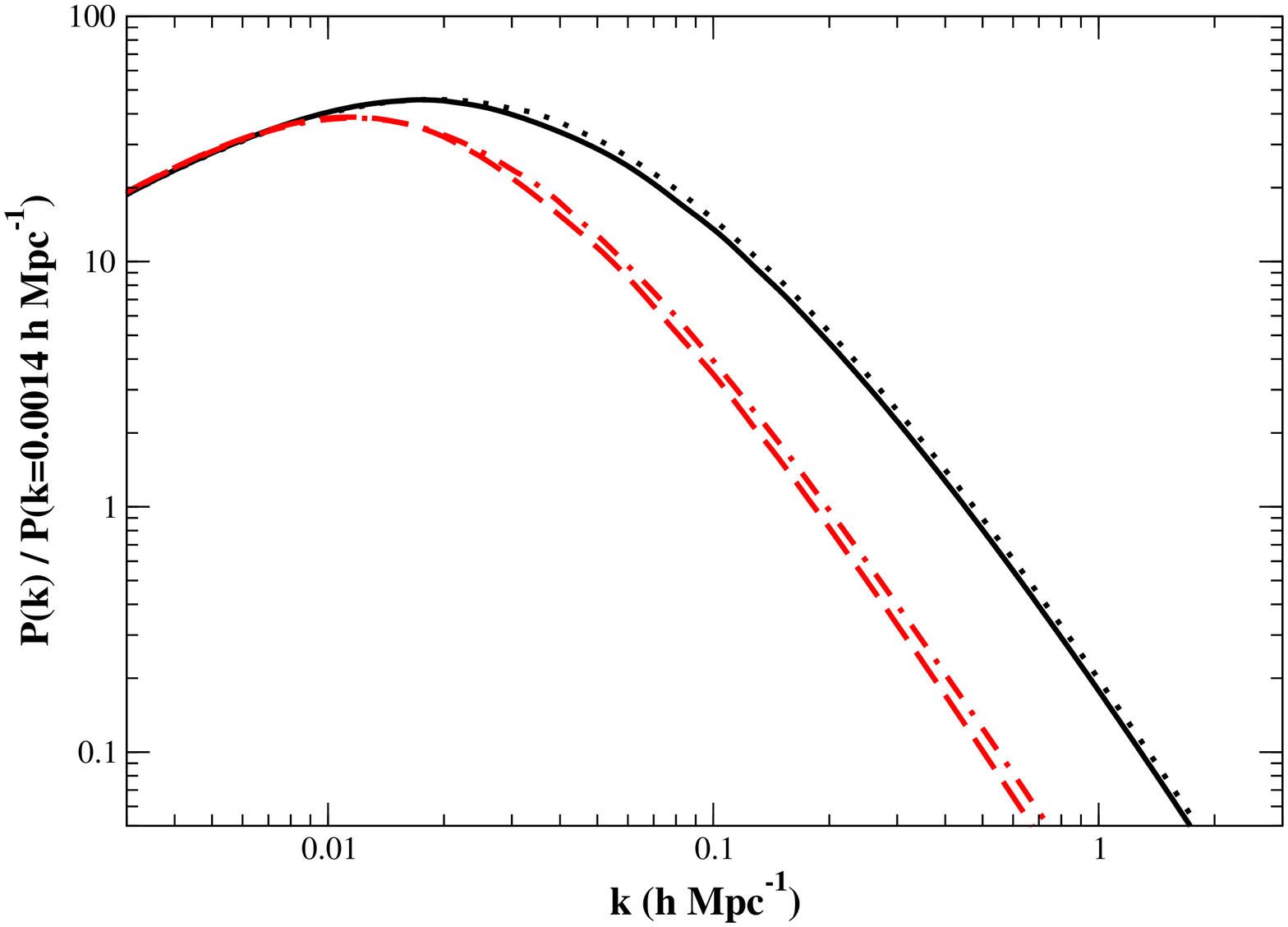}
\includegraphics[width=4cm,height=4cm,angle=-90]{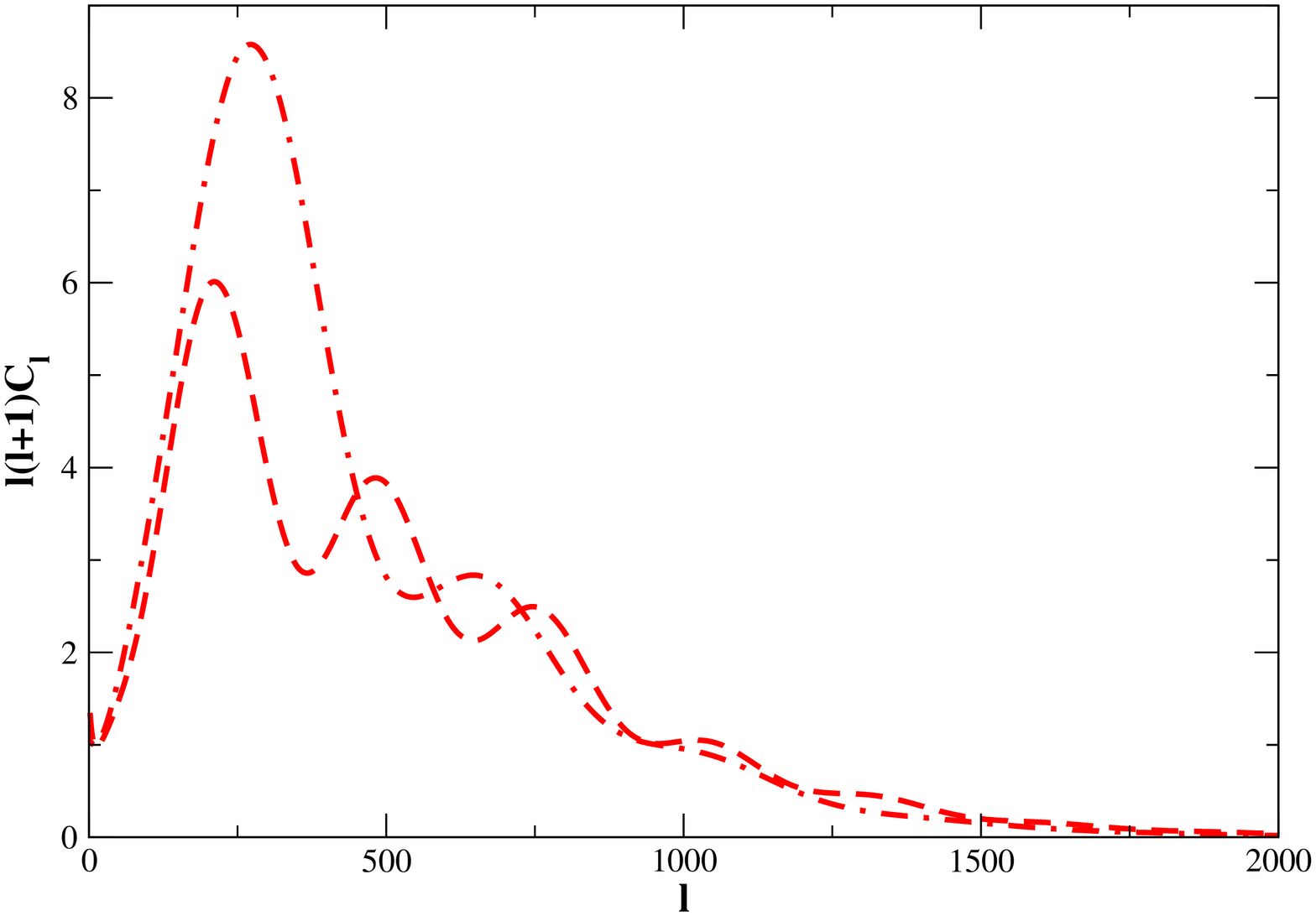}
\caption{\label{fig:fig1} Illustration of how combining data sets break degeneracies.  The left panel shows the matter power spectrum in four different 
models, where the baryon density has been artificially set low and a scale 
invariant power spectrum of adiabatic primordial perturbations has been 
assumed.   The 
black lines are for models with $\Omega_\nu = 0$.  The full line has 
$\Omega_{\rm m}=0.25$ and $h=0.735$, whereas the dotted line has $\Omega_{\rm m}=0.50$ and $h=0.3675$. The red lines are models with massive neutrinos.
The dashed line has $\Omega_{\rm m} = 0.25$, $\Omega_{\nu}=0.05$ and $h=0.735$, 
and the dot-dashed line has $\Omega_{\rm m}=0.50$, $\Omega_\nu = 0.10$ and 
$h=0.3675$.  Note that $f_\nu = 0.2$ in both cases.  The right panel 
shows the CMB power spectra for the two models with massive neutrinos, 
with the same line coding. The models have very different CMB power 
spectra, and hence measurements of the CMB anisotropies help constraining 
$\Omega _\nu$. }
\end{figure}

For scales below the neutrino free-streaming scale, $k> k_{\rm nr}$, 
the main effect is a reduction by a factor $1-f_\nu$ 
of the source term in the equation for the linear growth of density 
perturbation.  Hence, the amount of suppression is also set by the 
parameter $f_\nu$, and so it is this, and not $M_\nu$ which is 
directly accessible in measurements of large-scale structure.  
Since we can write $M_\nu = 93.14f_\nu \Omega_{\rm m}h^2\;{\rm eV}$,
we see that one also needs a constraint on $\Omega_{\rm m}h^2$ to 
go from a limit on $f_\nu$ to a limit on $M_\nu$.   Figure \ref{fig:fig1} 
illustrates this situation in a toy example.  
It is seen that the matter power spectrum must be combined with other probes, 
for example the CMB power spectrum, in order to provide stringent 
constraints on $M_\nu$.  

\section{Data and methods}

In our analysis we have used data from both CMB, observations of large
scale structures (LSS), type Ia supernovae (SNIa), baryon acoustic
oscillations (BAO), and additional priors on the Hubble parameter and
the baryon content of the universe. We have also applied constraints
on the cluster mass function from weak gravitational lensing. 

\subsection{CMB data}

The 3-year data release from the WMAP team \cite{hinshaw:2006} 
is at present the most constraining set of CMB observations.  The
WMAP experiment is a full-sky survey of CMB temperature anisotropies. With
this release, a Fortran 90 code for calculating the
likelihood of a given CMB power spectrum against the data was also provided. 
However, in ref. \cite{eriksen:2006} it was noted that this
likelihood code applied 
a sub-optimal likelihood approximation on large scales ($l \lesssim
30$), and in ref. \cite{huffenberger:2006} the authors pointed out an
over-subtraction of unresolved point sources on small angular scales ($l
\gtrsim 30$). Correcting for these two effects resulted in a slight
shift in some cosmological parameter values. In
ref. \cite{kristiansen:2006}, it was shown that these corrections to
the WMAP likelihood code also shifted the upper limit on the neutrino
masses from $M_\nu < 1.90$eV to $M_\nu < 1.57$eV when using WMAP data
only. In a revised version of the WMAP likelihood
code\footnote{http://lambda.gsfc.nasa.gov; version v2p2p1.}, these
corrections have to some extent been accounted for, and the results
using this code is in better agreement with refs. \cite{huffenberger:2006, kristiansen:2006} than the previous version. In
this paper we will make use of this revised likelihood code from the WMAP
team. We note, however, that the upper limits on $M_\nu$ would have been
slightly lower with the code used in refs. \cite{huffenberger:2006,
  kristiansen:2006}. 

In this work we are not considering CMB data from other experiments
than WMAP, since we want to restrict the number of data sets as much
as possible, and since these additional data sets have been shown not
to improve the limits on $M_\nu$ \cite{kristiansen:2006}. 

\subsection{Large scale structure}

Large scale structure surveys probe the matter distribution in the
universe by measuring the galaxy-galaxy power spectrum $P_g(k,z) =
\langle|\delta_g(k,z)|^2\rangle$. Since massive neutrinos give a
distinct imprint on the LSS due to their free-streaming effect, data
from galaxy surveys have proven to be important to put tight
constraints on neutrino masses. However, such surveys are also
troubled by bias effects that are hard to quantify. For example, it is
assumed that in the linear perturbation regime, the total matter power
spectrum, $P_m$, is proportional to the galaxy power spectrum by the
simple relation $P_g = b^2 P_m$. Recently, the use of this simple relation has
been debated \cite{percival:2006, smith:2006}, and it is at present
unclear what the precise corrections from a scale-dependent $b$
might be. 

There are two galaxy surveys of comparable size, the 2 degree Field
Galaxy Redshift Survey (2dFGRS) \cite{cole:2005}, and the Sloan
Digital Sky Survey (SDSS) main sample \cite{tegmark:2004}. As has been pointed
out in e.g. \cite{sanchez:2005, percival:2006, cole:2006}, there seems
to be some tension between the results from 2dFGRS
and SDSS. As a relevant example for this paper, the combination 
SDSS+WMAP prefers larger values of both $\Omega_{\rm m}$ and $\sigma_8$ (the 
rms mass fluctuations in spheres of radius $8 h^{-1}$ Mpc) than
what is found when using data from 2dFGRS+WMAP. In a recent analysis of 
the 2dFGRS-SDSS discrepancy \cite{cole:2006}, the authors conclude that it is caused by
non-linear bias effects. They also claim that SDSS is more sensitive
to these effects than the 2dFGRS sample, since the SDSS sample contains
more red galaxies, which are believed to cluster slightly different
from blue galaxies. 

Regardless of their origin, the reported inconsistencies between the 2dFGRS 
and SDSS galaxy surveys tell us that we should always be cautious when
including LSS data in our parameter estimation, at least until the
scale dependence of the bias parameter is better understood. 

When including 2dFGRS and SDSS data in our analysis we have tried to use
as commonly used analysis techniques as possible. We have therefore
stuck to the default analysis and parameter choices in the CosmoMC
code. For the 2dFGRS data this implies discarding all scales with $k>0.15
h\textrm{Mpc}^{-1}$. This corresponds to the choice made in
e.g. \cite{sanchez:2005}. For the 2dFGRS data we also use the prescribed correction for
nonlinearities given in \cite{cole:2005}, $P_g=(1+Qk^2)/(1+Ak)$ with
$A=1.4$ and $Q=4.0 \pm 1.5$. For SDSS we discard scales with
$k>0.20h\textrm{Mpc}^{-1}$, as is done in
e.g. \cite{tegmark:2003}. In this paper they find that using this cutoff scale yields results that are consistent with the what they find when applying more a more conservative cutoff on $k$.  No corrections for non-linearities are
included in our SDSS analysis. The treatment of cutoffs and
nonlinearities in the analysis of the 2dFGRS and SDSS data are
different. Thus, if any tension is found between these two data sets
here, it is not obvious whether this stems from problems within the
datasets or if it is caused by the way the data sets are analysed
here. However, any such tension would illustrate that there are
problems related to using LSS surveys for parameter estimation when
adapting these commonly used analysis methods for the two surveys. 

\subsection{Type Ia supernovae}

The luminosity distance-redshift relationship measured by observations 
of supernovae of type Ia (SNIa) provides the most direct evidence for 
cosmic acceleration and dark energy.  There are still open questions 
regarding both the exact mechanism behind these supernovae and their 
use as `standard candles', and so we choose not to rely on SNIa in 
our robust neutrino mass limit.  When we do use them, we use the 
 data from the Supernova Legacy Survey (SNLS) \cite{astier:2006}. 

\subsection{The cluster mass function}  

Massive galaxy clusters are extremely rare high-density peaks in the matter 
fluctuations, containing matter originating from a co-moving volume spanning 
$\sim 10 h^{-1}$ Mpc which has undergone gravitational collapse.  
Their abundances, as quantified by the CMF, are thus a sensitive probe of 
$\sigma_8$ and $\Omega_{\rm m}$. While all previous measurements of the CMF have been 
based on observations of baryonic tracers of cluster mass (with inherent uncertainties 
and possible biases which are not yet fully understood; see e.g.\ \cite{rasia:2006}), the CMF of \cite{dahle:2006} was derived from weak gravitational lensing measurements 
of the masses of a volume-limited sample of X-ray luminous clusters. 
The details of the data reduction and weak gravitational shear estimator are
given in \cite{dahle:2002}, and \cite{dahle:2006} describes the derivation of 
cluster masses from gravitational lensing measurements. The cluster sample of 
\cite{dahle:2006} was selected 
from a large volume of $8.0 \times 10^8 (h^{-1} {\rm Mpc})^3$ (assuming a spatially 
flat universe with $\Omega_{\rm m}=0.3$), above a threshold value in X-ray luminosity 
($L_{\rm X}$). 
Using $L_{\rm X}$ as a proxy for mass in the cluster selection could in principle 
introduce a baryonic bias to the CMF measurements, depending on $L_{\rm X}$ measurement 
uncertainties and the intrinsic
scatter around the mass-$L_{\rm X}$ relation. In \cite{dahle:2006}, any such effects 
were effectively removed, by carefully calibrating the amplitude of the mass-$L_{\rm X}$ 
relation and the scatter around the mean relation (thereby estimating the 
sample completeness as a function of mass), and by calculating cosmological parameter
constraints exclusively based on the clusters well above 
the mass threshold corresponding to the X-ray luminosity threshold of the sample 
(where the sample is virtually complete). Other statistical and systematic 
uncertainties in the derived cluster masses and the effects of these 
uncertainties on the CMF constraints are discussed in detail in \cite{dahle:2006}. 

Parameter constraints in the $\sigma_8 - \Omega_{\rm m}$ plane were obtained by 
fitting the observed cluster abundances in three mass intervals to  
theoretical predictions for the CMF \cite{sheth:1999}. 
  
We have found that the value of $\chi^2$ from the CMF from
\cite{dahle:2006} can be approximated by the fit-function
\bee
\chi^2_{\textrm{CMF}} = 10 000u^4 + 6726u^3 + 1230 u^2-4.09u + 0.004,
\ene
where $u = \sigma_8(\Omega_{\rm m}/0.3)^{0.37}-0.67$. This result is rather
 insensitive to the choice of theoretical mass function: As noted 
by \cite{dahle:2006}, replacing 
the CMF prediction used here \cite{sheth:1999} with a more recent predicted 
CMF based on 
N-body simulations \cite{warren:2006} (taking into account the differing mass 
definitions in these two works) only results in a shift of 0.01 in $u$.  

\subsection{Other priors}

We have also tested the sensitivity of neutrino mass limits on priors
from the Hubble Space Telescope Key Project on the Extragalactic Distance Scale (HST), 
and Big Bang nucleosyntesis (BBN) predictions.

From HST, we use a prior on the Hubble parameter of $h=0.72\pm0.08$
\cite{freedman:2001}, and from BBN we use a prior on the physical baryon
density today, $\Omega_b h^2 = 0.022 \pm 0.002$ \cite{burles:2001,
  cyburt:2004, serpico:2004}.  
There are hints of some tension between the WMAP constraint on 
$\Omega_{\rm b}h^2$ and the value of this quantity inferred from the 
$^{4}{\rm He}$ abundance \cite{steigman:2006}, so in our robust limit 
we will not use the BBN prior.  Since there is still some debate 
about the value of the Hubble constant derived from the HST Key Project 
\cite{blanchard:2003,sandage:2006}, we also drop this prior when deriving 
our robust limit. 

We have also added information on the position of the baryonic
acoustic oscillation (BAO) peak in the
luminous red galaxy (LRG) sample in the SDSS survey
\cite{eisenstein:2005}. This prior is implemented by an effective fit
function as described in \cite{goobar:2006}. Using an effective
parameter 
\bee
A_\textrm{BAO} = \left[ D_M(z)^2 \frac{z}{H(z)} \right]^{1/3}
\frac{\sqrt{\Omega_{\rm m} H_0^2}}{z}, 
\ene
where $D_M(z)$ is the comoving angular diameter distance, the authors impose
the constraint
\bee
A_\textrm{BAO} =  0.469 \left( \frac{n_s}{0.98} \right)^{-0.35} (1+0.94 f_\nu) \pm 0.017.
\ene  
Throughout our analysis we have also applied a top-hat prior on the
age of the universe, $10\textrm{Gyr} < \textrm{Age} < 20\textrm{Gyr}$. 

\subsection{Parameter estimation}

For the parameter estimations we have used the publicly available
Markov chain Monte Carlo code CosmoMC \cite{lewis:2002}. We have used
a basic seven-parameter model, varying the parameters $\{ \Omega_b h^2,
  \Omega_{\rm m}, \log{(10^{10}A_S)}, h, n_s, \tau, M_\nu \}$. $\Mnu$
  is defined in eq. (\ref{eq:omeganu}). For the other parameters, the exact
  definitions are given by the CosmoMC code. 
We have assumed
  spatial flatness, no running of the scalar spectral index, that
  the dark energy is a cosmological constant and that the tensor to
  scalar fluctuation amplitude is negligible. These assumptions are
  well motivated by current available data \cite{spergel:2006}. Also,
  adding these extra degrees of freedom do not affect the limits on
  $M_\nu$ drastically \cite{kristiansen:2006}. 

Often cosmological neutrino mass limits are found using a large number
of different data sets and priors simultaneously. In this analysis we
consider in total 48 different combinations of data sets and priors to
see how the different data sets alter the neutrino mass limit.

\section{Results} \label{sec:results}
 
The neutrino mass limits found in our analysis are summarized in Table
\ref{tab:Mnu}. The limits quoted in the table range from $\Mnu \lesssim
6$eV from using LSS data only, via $\Mnu < 1.75$eV from WMAP data
only, and down to $M_\nu < 0.40$eV for a combination of WMAP, SDSS, SNLS
BAO and HST data.

\begin{table}[htb]
\begin{tabular}{lcccc}
\hline \hline
Data set & no prior & CMF & HST & BBN \\
\hline
WMAP & {1.75eV} & {1.43eV} & {1.47eV}& {1.73eV}\\
WMAP+2dFGRS & {1.02eV} & {0.89eV}& {0.92eV}& {1.01eV}\\
WMAP+SDSS & {1.05eV} & {1.13eV}& {0.87eV}& {1.05eV}\\
WMAP+SNLS & {1.10eV} & {0.70eV}& {1.02eV}& {1.07eV}\\
SDSS & {5.8eV} & {4.6eV}& {6.1eV}& {6.0eV}\\
2dFGRS & {5.2eV} & {5.3eV}& {5.3eV}& {5.2eV}\\
WMAP+2dFGRS+SNLS & {0.81eV} & {0.64eV}& {0.76eV}& {0.80eV}\\
WMAP+SDSS+SNLS & {0.44eV} & {0.57eV}& {0.42eV}& {0.44eV}\\
WMAP+2dFGRS+SNLS+BAO& {0.84eV} & {0.69eV}& {0.79eV}& {0.84eV}\\
WMAP+SDSS+SNLS+BAO& {0.42eV} & {0.56eV}& {0.40eV}& {0.42eV}\\
WMAP+SDSS+BAO& {0.55eV} & {0.76eV}& {0.52eV}& {0.56eV}\\
WMAP+2dFGRS+BAO& {1.04eV} & {0.89eV}& {0.92eV}& {1.01eV}\\
\hline
\end{tabular}
\caption{Estimated 95\% C.L. upper limits on $M_\nu$. The CMF, HST and BBN
  priors are added one at the time to the combination of data sets
  given in the left column.}
\label{tab:Mnu}
\end{table}

\subsection{WMAP + priors}

As one might have expected, the inclusion of WMAP data is crucial for
finding a good neutrino mass limit. The upper limit from WMAP data
alone found here, $\Mnu < 1.75$eV, resides between the former results found by the
WMAP team for their 3 year data, $\Mnu < 2.0$eV \cite{spergel:2006},
and the results from ref. \cite{kristiansen:2006} using
the same data with a modified likelihood code, $\Mnu < 1.57$eV. In
this case we see that both a CMF and HST prior will help
significantly in constraining $\Mnu$, and adding the CMF prior to the
WMAP data yields an upper bound of $\Mnu < 1.43$eV. 

\begin{figure}
\begin{center}
  \epsfig{figure=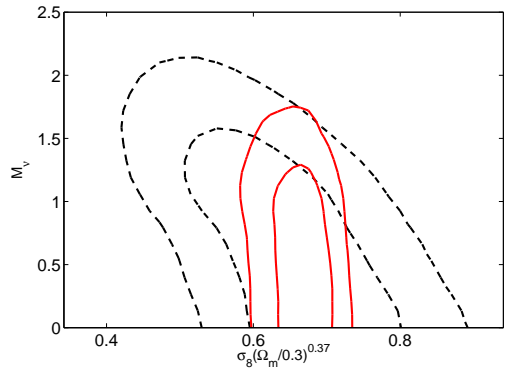,width=0.42\textwidth}
\caption{68\% and 95\% C.L. contours in the plane of $M_\nu$ and the
  effective parameter from the CMF constraint,
  $\sigma_8(\Omega_{\rm m}/0.3)^{0.37}$. The dashed black line shows the
  contours when using WMAP data only, and the solid black lines show the
results when adding the prior from the CMF.}
\label{fig:Mnu_and_eff}
\end{center}
\end{figure}

The CMF prior put
constraints on a combination of $\Omega_{\rm m}$ and $\sigma_8$, and in
Figure \ref{fig:Mnu_and_eff} we show confidence contours in the plane of
$\Mnu$ and $\sigma_8 (\Omega_{\rm m} /0.3)^{0.37}$ from WMAP data only, and
when we add the CMF prior. It is well known that there is a positive
correlation between $\Mnu$ and $\Omega_{\rm m}$ in the CMB power
spectrum. Both these parameters will alter the amplitudes of the CMB
peaks, by shifting the time of matter-radiation equality. At
this time neutrinos in this mass range were still relativistic, and
contributed to the radiation part. Keeping $\Omega_{\rm m}$ constant and
increasing $f_\nu$ will thus postpone the time of matter-radiation
equality and therefore enhance the amplitude of the acoustic peaks. To
shift the time of equality back and lower the acoustic peaks, one has
to increase $\Omega_{\rm m}$ correspondingly. However, the effective
parameter from the CMF constraint, $\hdp$, also contains the amplitude
parameter $\sigma_8$ which is negatively correlated with $M_\nu$ (from
the same reasoning). From Figure \ref{fig:Mnu_and_eff} we see that
this negative correlation is strong enough to make also the
correlation between $\Mnu$ and $\hdp$ negative; that is, small values
of $\hdp$ favor a large neutrino mass. When we add the CMF prior the
allowed region of low values of $\hdp$ shrinks, and the upper limit on
$\Mnu$ is reduced correspondingly. 

Note that with
this limit ($\Mnu < 1.43$eV) we have not used any LSS or SNIa data, or any other prior
on $H_0$ or $\Omega_b$. We therefore claim this neutrino mass
limit to be robust. 

\subsection{Including large scale structure}

\begin{figure}
\begin{center}
  \epsfig{figure=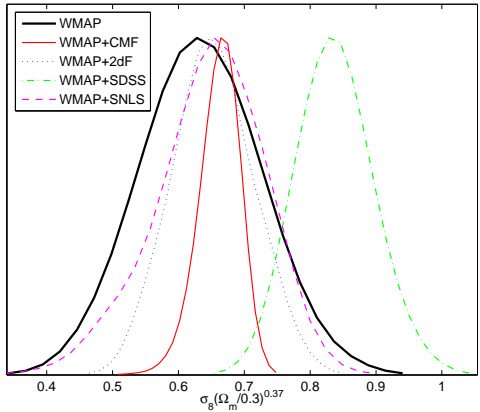,width=0.45\textwidth}
\caption{Marginalized one-dimensional distribution of $\sigma_8 \left(
    \Omega_{\rm m}/0.3 \right)^{0.37}$ when using WMAP data (thick black line),
  WMAP+CMF (solid red line), WMAP+2dFGRS (dotted blue line), WMAP+SDSS
  (dash-dotted green
  line) and WMAP+SNLS (dashed purple line).}
\label{fig:eff_1D}
\end{center}
\end{figure}

\begin{figure}
\includegraphics[width=4cm,height=4cm]{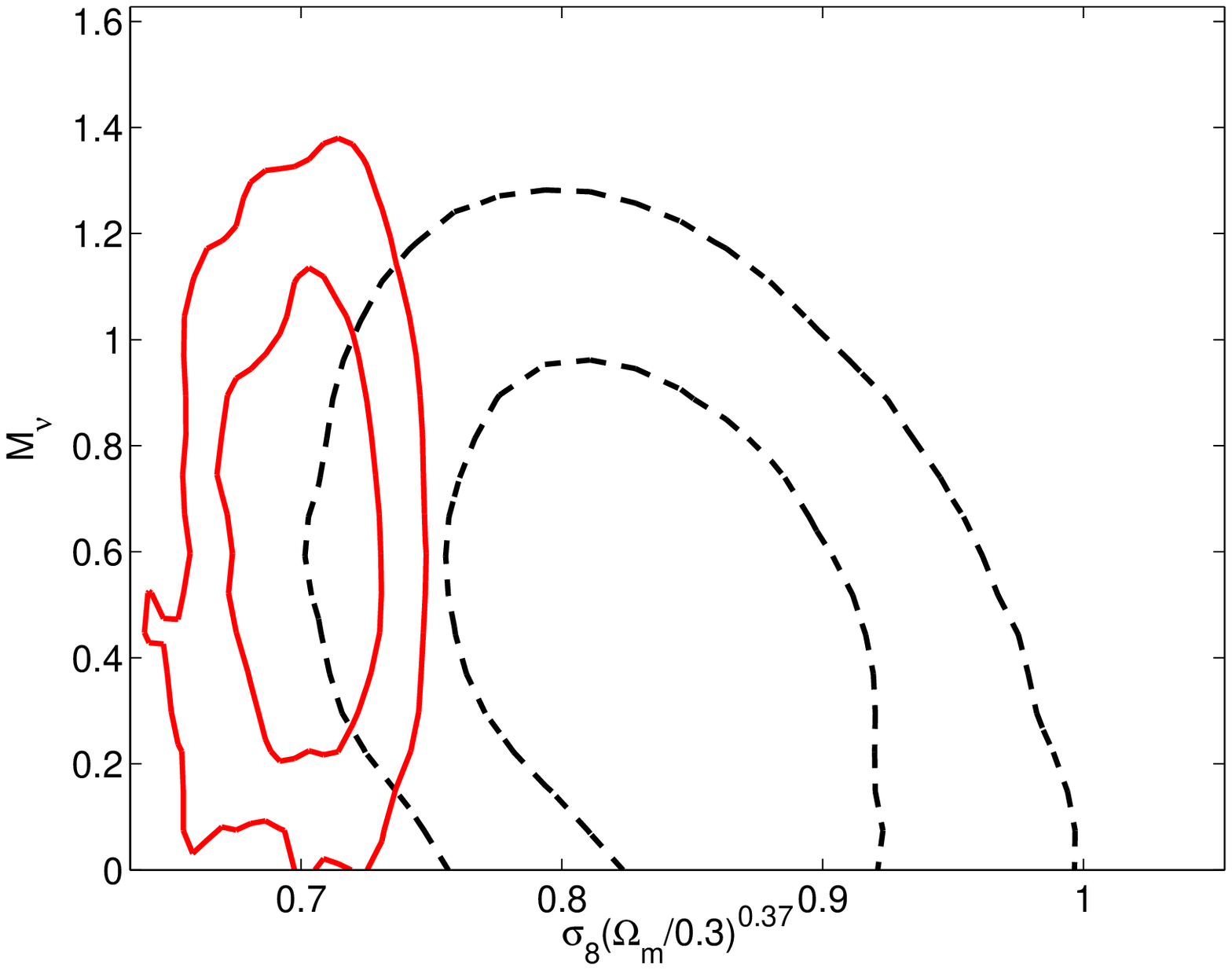} 
\includegraphics[width=4cm,height=4cm]{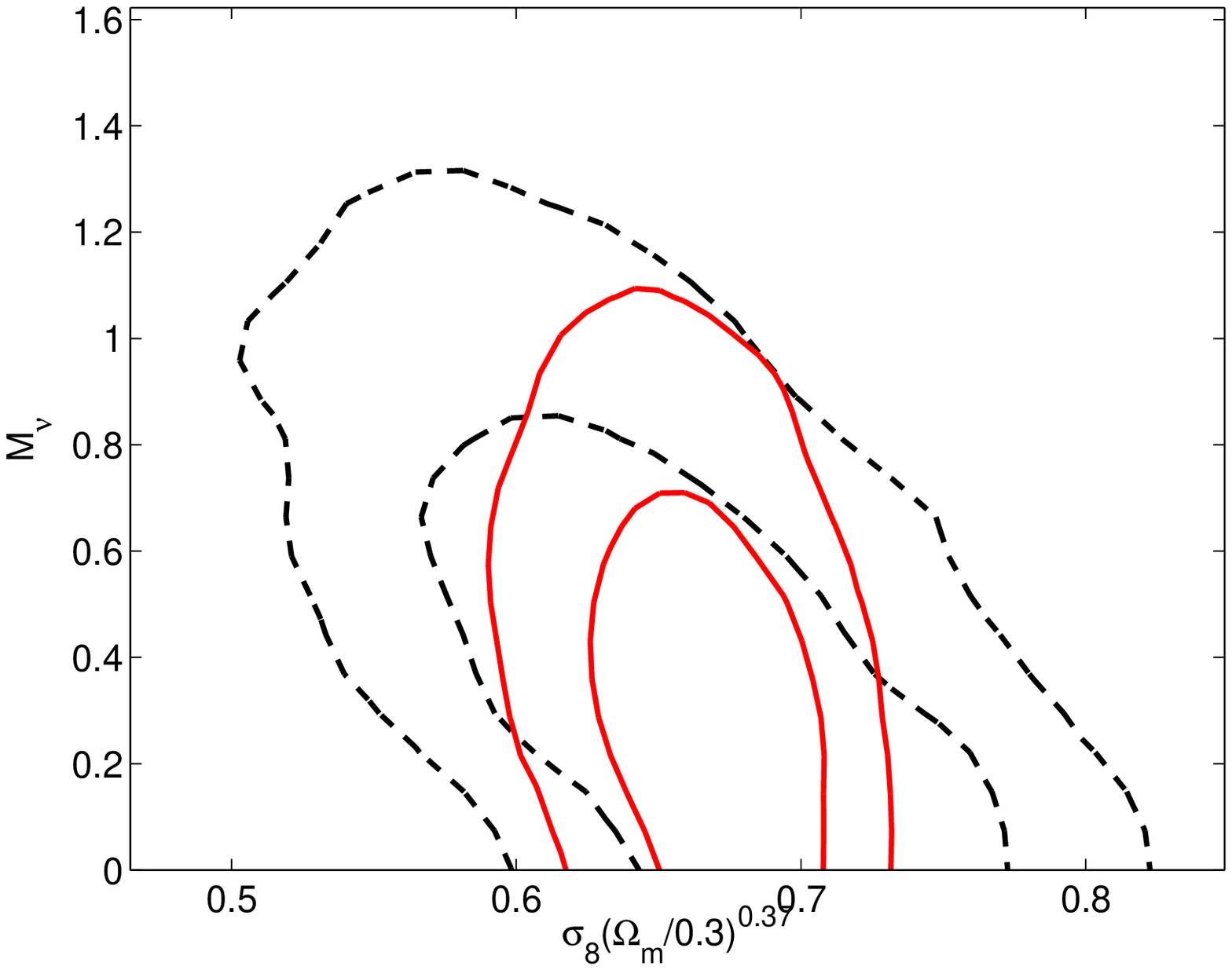} 
\caption{68\% and 95\% C.L. contours in the plane of $\hdp$ and
  $\Mnu$. Left panel: Contours from WMAP+SDSS (black, dashed line),
  and WMAP+SDSS+CMF (red, solid line). Right panel: Contours from
  WMAP+2dFGRS (black, dashed line), and WMAP+2dFGRS+CMF (red, solid line). }
\label{fig:Mnu_and_eff2}
\end{figure}

When adding LSS data the neutrino mass limit improves significantly,
giving $\Mnu < 1.02$eV for WMAP+2dFGRS and $\Mnu < 1.05$eV for WMAP
+SDSS. However, when adding the CMF prior, a strange effect
occurs. While this improves the $\Mnu$ limit to $\Mnu < 0.89$eV in the
case of WMAP+2dFGRS, the limit increases to $\Mnu <  1.13$eV with
WMAP+SDSS. This may indicate a inconsistency between the CMF prior and
our SDSS analysis. In Figure \ref{fig:eff_1D} we show the marginalized distribution
of the $\hdp$ parameter when using WMAP, WMAP+CMF, WMAP+2dFGRS, WMAP+SNLS or
WMAP+SDSS. For the latter, the distribution deviates significantly from
the four former, and we see that the WMAP+SDSS data set prefers larger
values of $\hdp$ than than the other combinations. This is in good
accordance with the results obtained in ref. \cite{sanchez:2005},
where they found that WMAP+SDSS preferred larger values for both
$\Omega_{\rm m}$ and $\sigma_8$ than what was found with WMAP alone or
WMAP+2dFGRS (using the WMAP 1-year data). Figure \ref{fig:eff_1D} indicates that we should be careful
when using data from SDSS in combination with 2dFGRS or CMF. Doing this,
the inconsistencies may produce artificially narrow parameter
distributions. In our case, the result of using WMAP+SDSS+CMF is that
we not only get an upper limit on $\Mnu$, but also a lower limit, such
that our 95\% C.L. limit becomes $0.23\textrm{eV}< \Mnu < 1.13$eV for
this combination of data sets. In Figure \ref{fig:Mnu_and_eff2} we
show the confidence contours in the plane of $\hdp$ and $\Mnu$ for
WMAP+2dFGRS and WMAP+SDSS with and without the CMF prior. From these
plots it is clear how the incompatibility of SDSS and CMF produces the
lower limit on $\Mnu$.

When adding LSS data to the WMAP data set, we see that the HST prior
is still important to improve the $\Mnu$ limits, while the BBN prior
is not needed. 

Also, it is interesting to note that adding the BAO prior to the
WMAP+2dFGRS data set combination has essentially no effect for the
neutrino mass limits, whereas adding the BAO prior to the WMAP+SDSS
data sets improves the neutrino mass limit by almost a factor
two. Again this illustrates the differences between the 2dFGRS and SDSS
sample in our analysis. The reason why the BAO prior is more important when using
SDSS, is the larger preferred value of $\Omega_{\rm m} h^2$ for the SDSS
sample. Adding the BAO prior will constrain the allowed region of
large $\Omega_{\rm m} h^2$, and thus also the allowed space for large $\Mnu$. 

We have also tried to constrain $\Mnu$ using LSS data alone, resulting
in mass limits of order $\Mnu \lesssim 6$eV. So although LSS data in
principle is a sensitive probe for neutrino masses, one needs to add
CMB data to break parameter degeneracies. 

\subsection{Including supernova data}

The inclusion of supernova data turns out to be as important as LSS
for constraining neutrino masses. Again, this can be understood by the
$\Mnu$-$\Omega_{\rm m}$ degeneracy. SNIa data is an effective probe of the
amount of dark energy in the universe, and under the flatness
assumption this will also automatically constrain $\Omega_{\rm m}$. In
Figure \ref{fig:eff_1D} we see that also the data set combination WMAP+SNLS
seems to be consistent with WMAP alone, WMAP+2dFGRS and CMF prior,
whereas it seems to be in some tension with our SDSS analysis. In the case of
WMAP+SNLS we see from Table \ref{tab:Mnu} that adding the CMF prior
improves the neutrino mass limit significantly, and that the inclusion of the CMF prior is a lot
more constraining for neutrino masses than both the HST and BBN prior
in this case. This means that if one believes in the SNIa
measurements,  we have an upper limit of $\Mnu<0.70$eV without using LSS
measurements at all.   

Note also that we can get a neutrino mass limit as low as $\Mnu <
0.40$eV without using Ly-$\alpha$ data, if we combine
WMAP+SNLS+SDSS+BAO+HST. But this limit weakens to $\Mnu < 0.79$eV by
substituting SDSS with 2dFGRS. Again, this illustrates how sensitive the
neutrino mass limit is to the choice of LSS sample and combination of priors.

\section{Conclusion} \label{sec:conclusion}

In this paper we have studied cosmological neutrino mass limits, and
how sensitive these limits are to different choices of data sets and
priors. We have also included a new prior from the cluster mass
function measured by weak gravitational lensing, which is a direct
probe of the total mass distribution in the universe. 

We report neutrino mass limits using 48 different combinations of data
sets and priors, and we find that the neutrino mass limit is very
sensitive to small changes in combinations of data sets and
priors. Especially striking results are found when interchanging data sets between the SDSS
and 2dFGRS galaxy surveys. For example will the combination WMAP+SDSS+BAO
give $\Mnu<0.52$eV at 95\% C.L., while WMAP+2dFGRS+BAO gives
$\Mnu<0.97$eV. These discrepancies occur because of a slight
inconsistency between the SDSS data as analysed here and many of the
other data sets used. Combining inconsistent data sets may of course lead to
unreliable results. Also, the tension between the two galaxy
surveys indicate that there may be systematics related to e.g. the
scale dependence of the bias parameter that have to be better
understood before we can rely fully on the LSS analysis.  
By discarding these data sets, and in addition refraining from using 
constraints from SNIa, HST, and BBN, we end up with a conservative, 
although robust cosmological
neutrino mass limit of $\Mnu < 1.43$eV from using only WMAP data in
combination with the CMF prior.

\acknowledgments

{\O}E, JRK and HD acknowledge support from the Research Council of
Norway through project numbers 159637, 162830 and 165491.
Some of the results in this work are based on observations made with the 
Nordic Optical Telescope, operated on the island of La Palma 
jointly by Denmark, Finland, Iceland,
Norway, and Sweden, in the Spanish Observatorio del Roque de los
Muchachos of the Instituto de Astrofisica de Canarias.



\end{document}